# Intercalation Leads to Inverse Layer Dependence of Friction on Chemically Doped MoS$_2$

Ogulcan Acikgoz,[1*] Enrique Guerrero,[2*] Alper Yanilmaz,[3*] Omur E. Dagdeviren,[4] Cem Çelebi,[3†] David A. Strubbe,[2†] and Mehmet Z. Baykara[1†]

[1]Department of Mechanical Engineering, University of California Merced, Merced, California 95343, USA

[2]Department of Physics, University of California Merced, Merced, California 95343, USA

[3]Department of Physics, Izmir Institute of Technology, Izmir 35430, Turkey

[4]Department of Mechanical Engineering, École de technologie supérieure, University of Quebec, Quebec H3C 1K3, Canada

* These authors contributed equally to this work.

† Corresponding authors: cemcelebi@iyte.edu.tr, dstrubbe@ucmerced.edu, mehmet.baykara@ucmerced.edu

We present results of atomic-force-microscopy-based friction measurements on Re-doped molybdenum disulfide (MoS$_2$). In stark contrast to the widespread observation of decreasing friction with increasing number of layers on two-dimensional (2D) materials, friction on Re-doped MoS$_2$ exhibits an anomalous, i.e., *inverse* dependence on the number of layers. Raman spectroscopy measurements combined with *ab initio* calculations reveal signatures of Re intercalation. Calculations suggest an increase in out-of-plane stiffness that inversely correlates with the number of layers as the physical mechanism behind this remarkable observation, revealing a distinctive regime of puckering for 2D materials.

Friction is among the most fascinating yet least understood subjects in classical mechanics. Despite its prevalence in mechanical systems, systematic studies aimed at uncovering the underlying physical mechanisms on fundamental length scales became only possible with the advent of modern experimental tools such as the atomic force microscope (AFM) [1, 2]. While such methods provide outstanding resolution in space and force, a comprehensive physical picture of frictional processes remains yet to be formed, mainly because the phenomenon is a complex function of the multi-scale structural, mechanical and chemical properties of the surfaces involved, as well as environmental factors such as temperature and humidity [3].

The discovery of the exotic electronic properties exhibited by graphene about fifteen years ago [4] and the ensuing boom in two-dimensional (2D) materials [5, 6] led to new avenues in fundamental friction research. In particular, the atomically smooth and chemically inert surfaces exposed by the majority of 2D materials provide a simplified platform on which AFM-based friction experiments can be performed [7]. Such studies also exhibit practical relevance, as 2D materials could potentially be employed as solid lubricants in micro- and nano-scale mechanical systems where surface-based phenomena such as friction and wear bear increasing importance, and conventional, fluid-based lubrication schemes are not feasible [8-10].

A particularly crucial discovery in AFM-based 2D material friction research is that the friction decreases with increasing number of layers, as first reported in milestone experiments by Filleter *et al.* [11] and Lee *et al.* [12]. These observations were later confirmed by a number of independent studies performed on graphene, molybdenum disulfide ($MoS_2$), and other 2D materials [13-15], thus establishing decreasing friction with increasing number of layers as a seemingly universal characteristic of 2D materials. Among various theories proposed to explain the underlying physical mechanisms, the one that gained the most traction in the literature is the *puckering* effect [12]. In particular, it is proposed that the sharp AFM tip sliding on a 2D material leads to the formation of a *pucker* (bulge) ahead of the tip, thus leading to an increase in contact area and enhanced friction. As the number of layers increases, the sample's bending stiffness increases as it would for a thicker beam, resulting in the suppression of the pucker and thus decreasing friction [16] (bending stiffness is also determined by in-plane stiffness, but that has little dependence on the number of layers in 2D materials [17, 18]). Puckering is also reduced by increased adhesion to a substrate which hinders out-of-plane deformation, and thus reduces the layer dependence [19]. While the idea of puckering, which has been studied in detail via computational approaches [16], can explain the trend of decreasing friction with increasing number of layers, other theories have also been proposed to explain the trend, ranging from a suppression of electron-phonon coupling (and thus a reduction in energy dissipation) [11] to decreasing surface roughness [13] with increasing number of layers. On the other hand, no direct studies have been conducted to address the question of whether this apparently ubiquitous layer-dependence trend of friction on 2D materials can be suppressed or even reversed through certain approaches, including but not limited to the application of strain [20], electrostatic fields [21], and chemical doping [22].

Motivated as above, we report here, by way of AFM-based friction measurements performed on rhenium (Re)-doped $MoS_2$, the observation of *inverse* layer dependence of friction, in stark contrast to the seemingly universal trend of decreasing friction with increasing number of layers of 2D materials. Raman spectroscopy measurements interpreted by *ab initio* density functional theory (DFT) calculations indicate that the Re dopants are intercalated between $MoS_2$ layers. DFT additionally reveals that Re intercalants lead to an increase in out-of-plane stiffness, an effect that decreases with increasing number of layers, but that the dopants do not significantly affect the interaction of an AFM tip with a rigid, flat $MoS_2$ surface. Further details and analysis of the calculations are presented in a forthcoming paper [23].

We started our investigation by studying the layer dependence of friction on undoped MoS$_2$. As demonstrated in Fig. 1(a,b) for a stair-like flake that progressively features one-to five-layer regions, the layer-dependence results obtained via friction force maps recorded on undoped MoS$_2$ are in harmony with previous experimental studies in the literature [12]. In particular, the friction force is monotonically decreasing with increasing number of layers, pointing toward an enhanced solid lubrication effect with increasing thickness.

Unlike undoped MoS$_2$, AFM-based friction measurements on Re-doped MoS$_2$ reveal that Re-doped flakes exhibit a completely unexpected inverse layer dependence of friction. In particular, results reported in Fig. 1(c,d) for a Re-doped MoS$_2$ flake with one-, two-, and three-layer regions show a striking contrast to those in Fig. 1(a,b). Specifically, while solid lubrication is still achieved with Re-doped MoS$_2$ (i.e. the friction force recorded on Re-doped MoS$_2$ is always lower than the underlying SiO$_2$ substrate), single-layer Re-doped MoS$_2$ exhibits the lowest friction force and the friction force increases with the number of layers, in violation of the seemingly universal rule of decreasing friction with increasing number of layers.

To confirm the anomalous results obtained on Re-doped MoS$_2$ and ensure that the findings are not specific to one flake, measurements were repeated on a different Re-doped MoS$_2$ flake with two-, eleven-, thirteen-, fourteen- and fifteen-layer regions. The results (Fig. 2) demonstrate a similar overall trend: increasing friction with number of layers, with the trend reaching an apparent saturation after fourteen layers. By comparing to friction on the SiO$_2$ substrate, we find that Re-doped MoS$_2$ exhibits generally lower friction than pristine MoS$_2$, and the highest number of layers have friction comparable to pristine MoS$_2$ [24].

It needs to be pointed out that an increasing friction trend with increasing number of layers was shown once before, on undoped MoS$_2$ samples [15], and attributed to an exceptionally large AFM probe apex. In particular, the large apex led to a weakening of the puckering effect such that the layer dependence trend is now dominated by the corrugation of the potential-energy landscape experienced by the probe apex as it slides over different samples. On the other hand, the "regular" results obtained on undoped MoS$_2$ using the same AFM probe in our experiments (Fig. 1(a,b)) exclude a possible link between probe characteristics and the unusual findings on Re-doped MoS$_2$. It also needs to be emphasized that the measurements were repeated on multiple undoped and Re-doped samples on different days, with the same trends observed on both types of samples. The sharpness of the step edges in the friction maps provide further proof for the absence of an exceptionally blunt apex. Based on these observations, we are confident that the observed anomalous trend is intrinsic to Re-doped MoS$_2$ and not probe-dependent.

To explore the physical reasons behind the observation of an inverse layer-dependence of friction on Re-doped MoS$_2$, we first checked the potential presence of unexpected trends in adhesion and roughness with increasing number of layers. The results of these investigations do not yield any significant trends in the layer-dependent behavior of adhesion and roughness that would explain the anomalous trend in friction we observe for the Re-doped samples [24]. Subsequently, we performed Raman spectroscopy measurements on single-layer, few-layer (i.e., with less than 10 layers) and bulk flakes of undoped and Re-doped MoS$_2$, the results of which are summarized in Fig. 3. The main conclusions from these measurements can be described as follows: (i) the separation in frequency between the E$_{2g}$ and A$_{1g}$ modes (19 cm$^{-1}$, 23 cm$^{-1}$, and 25 cm$^{-1}$ for single-layer, few-layer and bulk MoS$_2$ regions, respectively) decreases with decreasing thickness in accordance with the literature [25]; (ii) the spectra of Re-doped MoS$_2$ are devoid of peaks associated with the formation of ReS$_2$ (an E$_{2g}$ peak at 163 cm$^{-1}$ and an A$_{1g}$-like peak at 213 cm$^{-1}$), ruling out phase segregation as a result of Re doping [26]; (iii) there is a significant decrease in the intensity of the E$_{2g}$ and A$_{1g}$ peaks for all Re-doped samples (calibrated against the intensity of the reference Si peak at 521 cm$^{-1}$) when compared with the undoped ones; and (iv) the two predominant Raman-active modes of MoS$_2$ (the E$_{2g}$ mode, which arises from the in-plane opposite vibration of two S atoms against a Mo atom, and the $A_{1g}$ mode, which corresponds to the out-of-plane vibrations of S atoms in opposite directions) [27] are observed in all samples, with small red shifts in the few- and many-layer cases.

The observations about intensity changes (iii) and peak shifts (iv) allow us to infer the doping site for Re. For transition-metal dichalcogenides, particularly in many-layer form, determination of doping site is a significant experimental challenge, only definitively resolved in a few cases [28, 29]. Intercalation between the layers, and substitution for the chemically similar Mo, are the most typical sites for a transition-metal dopant such as Re in $MoS_2$. We performed plane-wave DFT calculations with Quantum ESPRESSO [30, 31] of Raman spectra [32] (see further methodological details in [24]). We studied bulk (multi-layer) $MoS_2$, undoped and with Re in tetrahedral (t-) intercalation and Mo substitution sites, to identify frequency shifts in the two prominent peaks (Fig. 4). t-intercalation has a red shift in both peaks, consistent with the measurements in Fig. 3(c), whereas Mo substitution has a blue shift for $E_{2g}$, thus pointing to t-intercalation as the predominant dopant site. This calculated blue shift matches Raman measurements by Gao *et al.* for a monolayer $MoS_2$ sample believed to have Mo substitution on the basis of electron microscopy and photoemission [33]. Apart from the peak shifts, the significant decrease in peak intensities relative to undoped $MoS_2$ is consistent with similar observations reported before for $MoS_2$ intercalated with Co [34] and Li [35]. This decrease was also seen for intercalation into other 2D materials such as $MoSe_2$, $WSe_2$, $WS_2$, and graphene [34]. Our DFT calculations point to a decrease but of lesser magnitude for t-intercalation, but also for Mo-substituted $MoS_2$; and little effect was seen in DFT calculations of Ni-intercalated $MoS_2$ [36]. As a result, the mechanism for the experimental observations is unclear and may relate to dynamical effects in the Raman tensor. Regardless of mechanism, the decreased Raman intensities together with the peak redshifts indicate that the Re dopants are primarily intercalated in the sample. We note that two layers at minimum are required for true intercalation, and therefore the Re-doped one-layer samples studied here may have Re atoms between the MoS$_2$ layer and the SiO$_2$ substrate, which presumably increases adhesion.

Now that we have determined that Re dopants primarily intercalate between MoS$_2$ layers, we turn our attention to establishing a physical picture that would explain the inverse layer dependence of friction observed on such samples. In order to achieve this goal, we performed DFT calculations of out-of-plane stiffness (i.e. elastic parameter $C_{33}$) for bulk structures of undoped (i.e. pristine) MoS$_2$ as well as Re-doped MoS$_2$ in the t-intercalation and Mo substitution cases [24]. The results of the calculations, summarized in Fig. 5(a) for supercells of decreasing size (which correspond to increasing Re concentration), show an overall increase in out-of-plane-stiffness for Re-doped MoS$_2$ when compared with the pristine material. This effect is significantly more pronounced for t-intercalation than Mo-substitution, due to the formation of interlayer covalent bonds by Re, similar to the general increase in interlayer coupling for Ni-intercalated MoS$_2$ [37]. More importantly for the present discussion, the stiffening effect is proportional to the dopant concentration, suggesting a model of rigid layers connected in series by springs, in which the dopant contributes a stronger spring constant (Fig. 5(a), inset). In other words, the stiffening effect induced by a fixed number of Re dopants will be less pronounced for a larger number of layers. As revealed by AFM simulations [16], there is a small volume around the AFM tip which is elastically deformed, and the local elastic properties control the degree of puckering. For the low doping in our experiments (estimated as 0.1%), the number of Re dopants in this volume is almost always 0 or 1, and therefore the local dopant concentration is lower for more layers, resulting in a weaker stiffening effect.

While previous discussions of puckering in 2D materials have focused on the interplay of bending stiffness and substrate adhesion [16, 19] it is implicit in these studies that out-of-plane deformation in the presence of substrate adhesion is also hindered by the out-of-plane stiffness. For multi-layers, this out-of-plane stiffness is similar to adhesion to the underlying layers, which can be modeled elastically [38], and is also related to a binding energy between layers [16]. While a comprehensive model has not been established for the interplay of bending, out-of-plane stiffness, and substrate adhesion in determining puckering, it is clear that as out-of-plane stiffness increases there can be a crossover from the usual regime dominated by bending stiffness to one dominated by out-of-plane stiffness, just as a substrate-adhesion-dominated regime has been explored [19]. Our Re-doped $MoS_2$ samples are in this distinctive regime, as depicted in Fig. 5(b-e). Since out-of-plane stiffness reduces puckering, and for a given number of dopants the increased stiffness decreases with the number of layers, more puckering and consequently, a higher value of friction will be observed with increasing number of layers, consistent with the experiments. Two key implications of our model which are seen in the data are: friction on Re-doped $MoS_2$ is generally lower than on pristine $MoS_2$, and many-layer Re-doped $MoS_2$ friction reaches a limit similar to pristine $MoS_2$ [24]. Our model also suggests greater variation in friction between different regions will be found for Re-doped $MoS_2$ compared to pristine, depending on whether a dopant is present nearby, but such an effect has not been resolved in the experiments. Additionally, the model suggests that as dopant concentration increases, and the local dopant concentration ceases to depend on number of layers, the layer dependence will be reduced or even return to the pristine trend, determined by layer stiffening rather than dopant stiffening; this can be a possible target of future experiments. Finally, we note that simulations have suggested that—even in the absence of puckering—out-of-plane elasticity can be a determining factor in friction on graphite [39], so there could be additional mechanisms by which out-of-plane stiffness contributes to the observed layer dependence.

For comparison to this latter explanation, we considered the possibility that the friction trends are due to changes in the potential-energy landscape. We used DFT to evaluate the friction forces that would be experienced by a model AFM tip apex [40] sliding on Re-doped

MoS$_2$ as a function of the number of layers, in the absence of puckering [24]. We find that friction forces *decrease* as the first few layers are added and then saturate, opposite to the experimental dependence for few layers, and having no effect for many layers unlike the experiments. These findings are at significant variance with the experiments and thereby leave alterations to the degree of puckering (which overwhelm the effect of the potential-energy landscape) as the most likely explanation for the experimental observations.

We presented here the remarkable observation of an inverse layer dependence of friction on Re-doped MoS$_2$, in violation of the general understanding that friction on 2D materials decreases with increasing number of layers. Informed by Raman spectroscopy measurements and *ab initio* calculations, we proposed a mechanism of decreasing out-of-plane stiffness with increasing number of layers for a given number of intercalated dopants, which leads to an enhanced effect of puckering with increasing sample thickness and consequently, higher friction. Our results indicate the presence of a distinct regime of puckering where out-of-plane stiffness rather than bending stiffness or substrate adhesion is the decisive factor. This study opens the way for selective tuning of friction in micro- and nano-scale mechanical systems, by the combined use of undoped 2D materials and those with intercalated dopants (which are most probably not limited to the Re-doped MoS$_2$ system investigated here). On the other hand, more work will need to be conducted to determine if there is a limit to the friction increase with increasing number of layers (as suggested by the data presented in Fig. 2, and our model) and the effect of doping concentration.

**Acknowledgments**

This work was supported by the Merced nAnomaterials Center for Energy and Sensing (MACES) via the National Aeronautics and Space Administration (NASA) Grant Nos. NNX15AQ01 and NNH18ZHA008CMIROG6R. We thank Sefaattin Tongay for graciously providing the Re-doped $MoS_2$ samples. O.E.D. acknowledges Canada Economic Development Fund, Natural Sciences and Engineering Research Council of Canada, and Le Fonds de Recherche du Québec - Nature et Technologies. Computational resources were provided by the Multi-Environment Computer for Exploration and Discovery (MERCED) cluster at UC Merced, funded by National Science Foundation Grant No. ACI-1429783, and by the National Energy Research Scientific Computing Center (NERSC), a U.S. Department of Energy Office of Science User Facility operated under Contract No. DE-AC02-05CH11231.

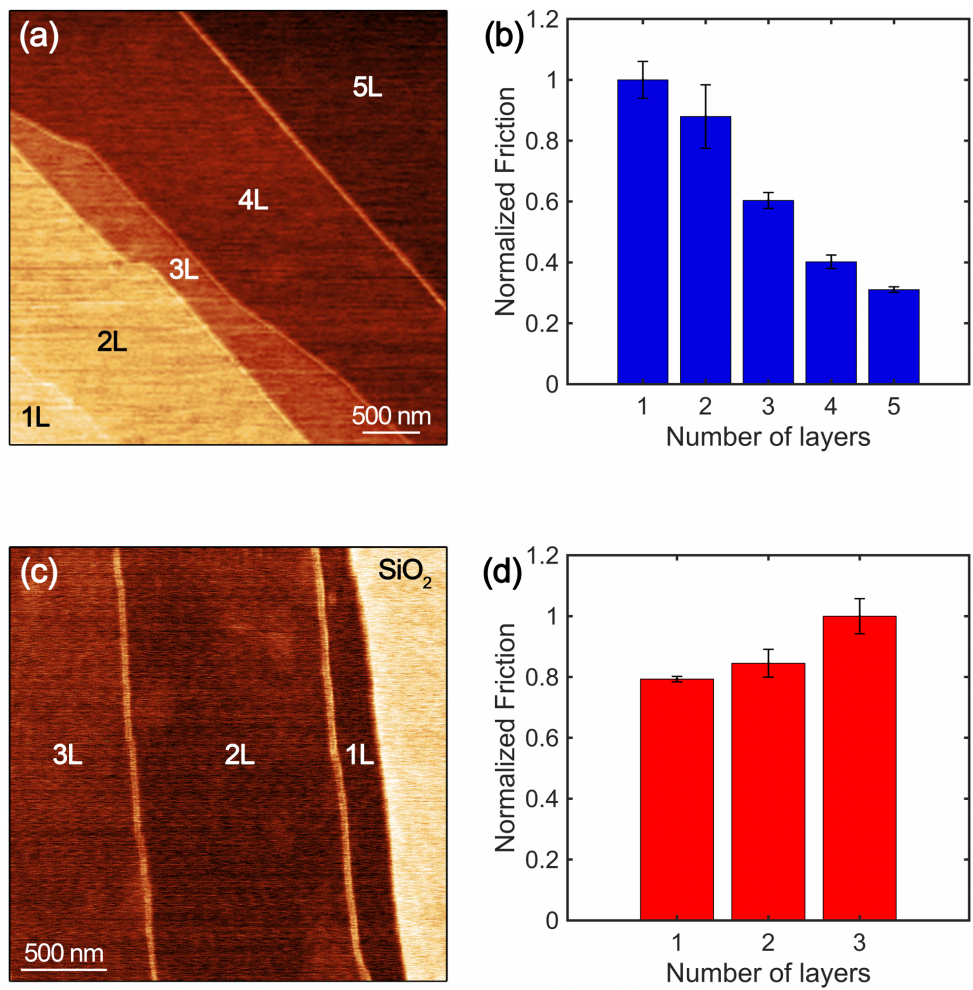

**FIG. 1.** (a) Friction force map obtained on an undoped MoS$_2$ flake with 1, 2, 3, 4, and 5 layers (1L, 2L, 3L, 4L, and 5L, respectively) situated on a SiO$_2$ substrate. (b) Friction on undoped MoS$_2$ areas with different number of layers. Friction is normalized to the value obtained on the 1L area. (c) Friction force map obtained on a Re-doped MoS$_2$ flake with 1, 2, and 3 layers (1L, 2L, and 3L, respectively), situated on a SiO$_2$ substrate. (d) Friction on Re-doped MoS$_2$ areas with different number of layers. Friction is normalized to the value obtained on the 3L area.

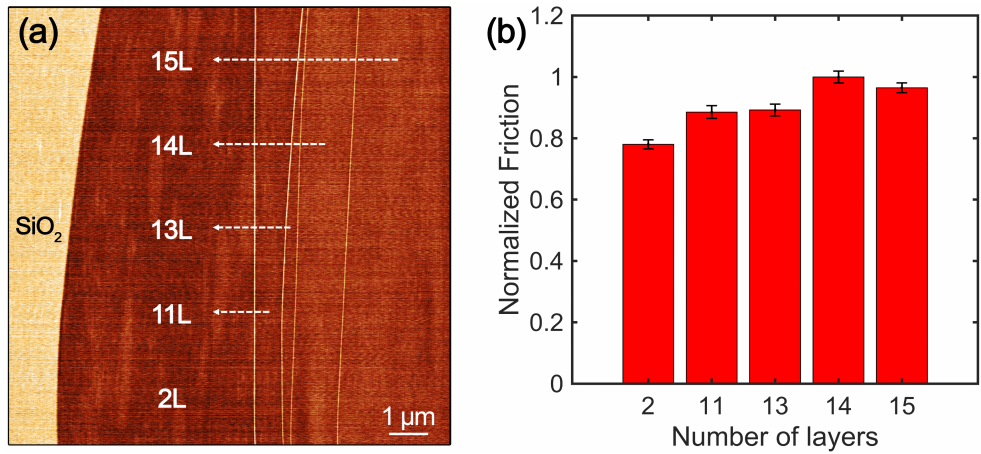

**FIG. 2.** (a) Friction force map obtained on a Re-doped $MoS_2$ flake with 2, 11, 13, 14, and 15 layers (2L, 11L, 13L, 14L, and 15L, respectively), situated on a $SiO_2$ substrate. (b) Friction on Re-doped $MoS_2$ areas with different number of layers. Friction is normalized to the value obtained on the 14L area.

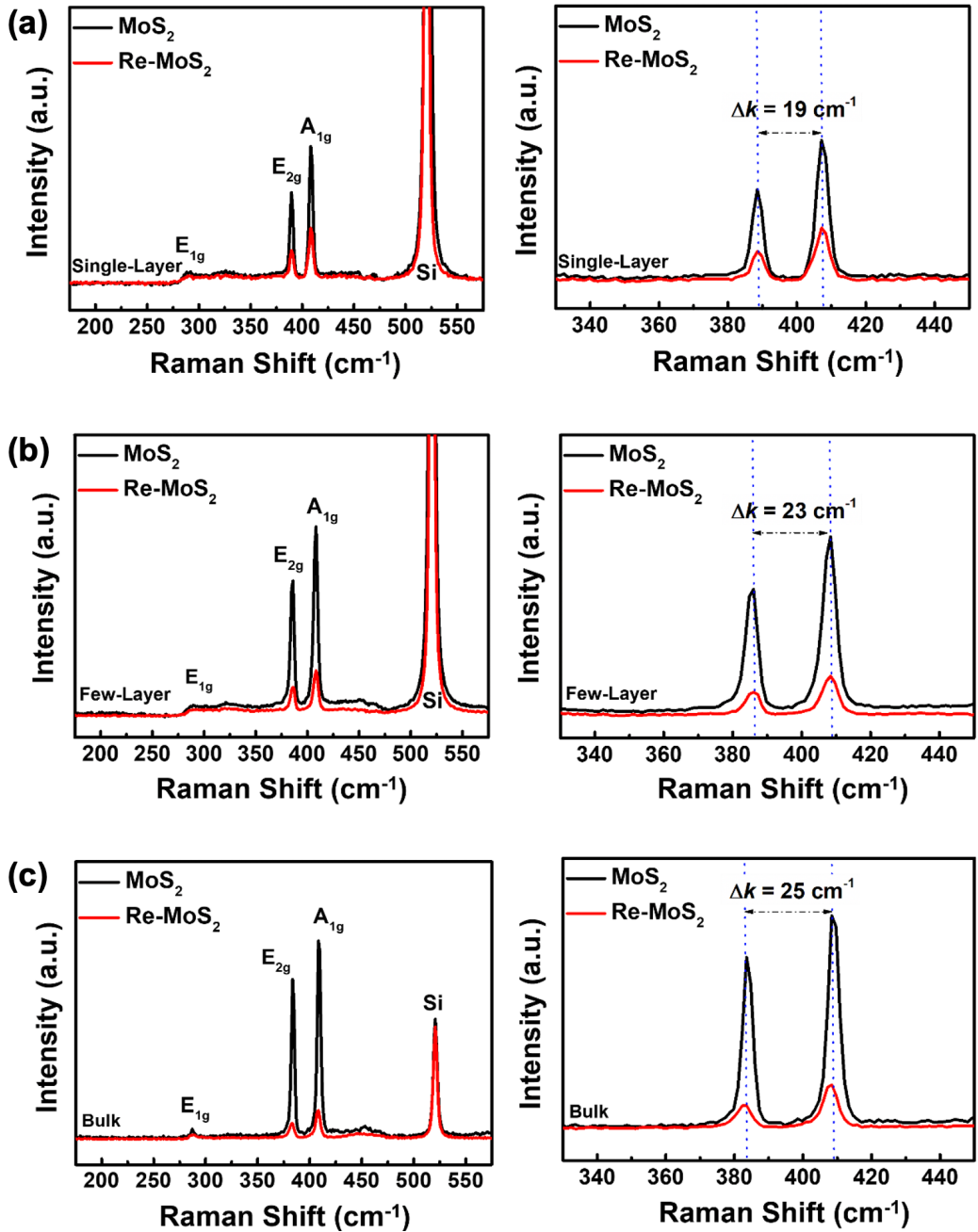

**FIG. 3.** Raman spectra of (a) single-layer, (b) few-layer and (c) bulk samples of undoped and Re-doped MoS$_2$. The panels on the right are zooms on the regions that contain the E$_{2g}$ and A$_{1g}$ peaks of MoS$_2$. Δ$k$ indicates the wavenumber spacing between the E$_{2g}$ and A$_{1g}$ peak positions, which are themselves highlighted by the dotted blue lines.

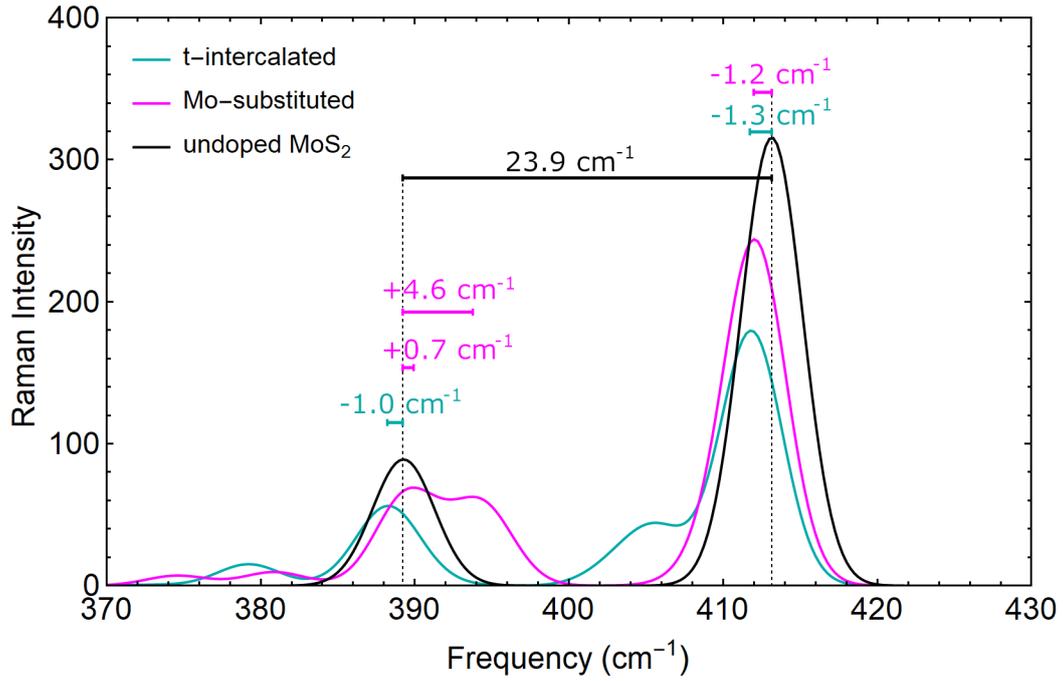

**FIG. 4.** DFT-calculated Raman spectra for bulk (many-layer) undoped and Re-doped MoS$_2$ structures, using 2×2×1 (in-plane) supercells for doped MoS$_2$. The red shifts of E$_{2g}$ and A$_{1g}$ for t-intercalation best match the experimental measurements of few- and many-layer Re-doped MoS$_2$. Raman intensities are in units of D$^2$/Å$^2$-amu, per MoS$_2$ unit, and a Gaussian broadening of 2 cm$^{-1}$ is used.

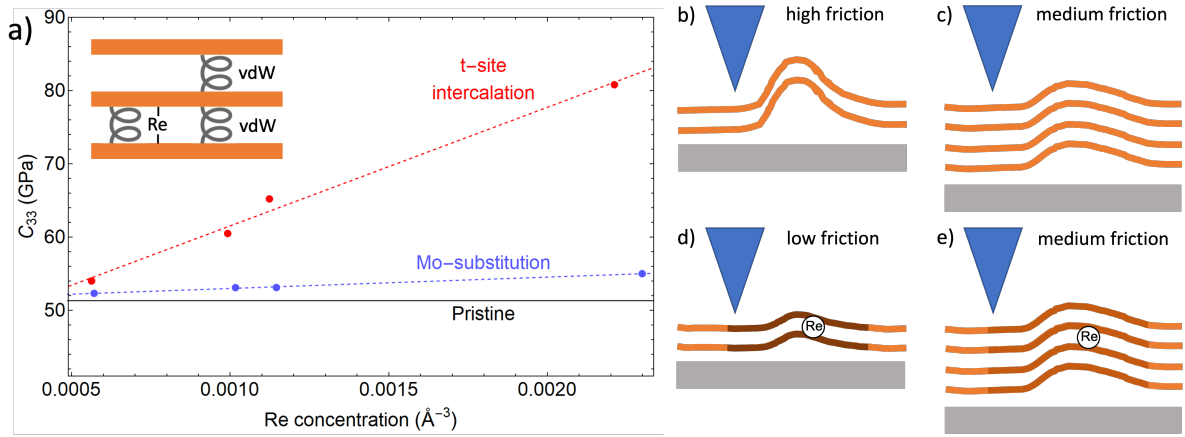

**FIG. 5.** (a) The out-of-plane elasticity coefficient ($C_{33}$) for bulk Re-doped MoS$_2$ as a function of concentration for several different supercells with one Re per cell, showing stiffening compared to the pristine material. Inset explains the linear dependence of $C_{33}$ on dopant concentration through a basic model of MoS$_2$ sheets elastically coupled by van der Waals forces, together with a single intercalated Re dopant acting as an additional, stronger "spring" due to interlayer covalent bonds. The effect of Mo-substituted Re on $C_{33}$ is much less than that for intercalated Re. (b-d) Schematic explanation of friction trends in terms of puckering, modulated by bending stiffness (increasing with number of layers) and out-of-plane stiffness (enhanced by the presence of a Re dopant but to a smaller extent with increasing number of layers). The blue triangle is the AFM tip, the gray rectangle is the substrate, and the orange lines are MoS$_2$ layers, with regions stiffened by Re colored darker in proportion to the effect, containing typically at most one Re atom. (b) undoped, few layers; (c) undoped, many layers; (d) doped, few layers; (e) doped, many layers.

# Supplementary Material for:

# Intercalation Leads to Inverse Layer Dependence of Friction on Chemically Doped $MoS_2$

Ogulcan Acikgoz,[1*] Enrique Guerrero,[2*] Alper Yanilmaz,[3*] Omur E. Dagdeviren,[4] Cem Çelebi,[3†]

David A. Strubbe,[2†] and Mehmet Z. Baykara[1†]

[1]Department of Mechanical Engineering, University of California Merced, Merced, California 95343, USA

[2]Department of Physics, University of California Merced, Merced, California 95343, USA

[3]Department of Physics, Izmir Institute of Technology, Izmir 35430, Turkey

[4]Department of Mechanical Engineering, École de technologie supérieure, University of Quebec, Quebec H3C 1K3, Canada

* These authors contributed equally to this work.

† Corresponding authors: cemcelebi@iyte.edu.tr, dstrubbe@ucmerced.edu, mehmet.baykara@ucmerced.edu

## 1. Experimental Methods and Results

### *1.1 Sample preparation*

Both the undoped and Re-doped $MoS_2$ crystals (at a doping concentration of 0.1%) studied here were synthesized over several weeks using the chemical vapor transport technique in quartz ampoules at high temperatures, resulting in bulk crystals of 1 mm and above in size. Prior to the AFM and Raman spectroscopy experiments, flakes of Re-doped and undoped $MoS_2$ were deposited onto $SiO_2$ substrates via mechanical exfoliation.

### *1.2 Atomic force microscopy*

The AFM measurements were performed under ordinary laboratory conditions (Room temperature: 22 – 23 °C; Relative humidity: 20 – 40%) using a commercial AFM instrument (*Asylum Research, Cypher VRS*). Both Re-doped and undoped $MoS_2$ samples were characterized with the same AFM probe (*Nanosensors, CDT-CONTR*) with a normal spring constant value $k$ of 0.84 N/m as determined by the Sader method [1]. The number of layers were determined from AFM topography maps. Friction force maps were recorded using established procedures [2].

*1.3 Raman spectroscopy*

Raman spectroscopy measurements were conducted using a commercial system (*Princeton Instruments, TriVista CRS*). Raman signals were recorded in a spectral range between 100 cm$^{-1}$ and 600 cm$^{-1}$ using an Ar$^+$ ion laser with a 514 nm excitation (600 grooves/mm grating) wavelength. For each sample, Raman measurements were repeated several times at different locations to ensure reproducibility. Each spectrum was normalized using *TriVista* software. The Si peak at 521 cm$^{-1}$ was used as a calibration reference. The Raman spectra on undoped and Re-doped MoS$_2$ samples were recorded concurrently. As both types of samples were synthesized around the same time, this resulted in very similar air exposure times.

*1.4 Roughness Measurements*

To try to identify the physical mechanism responsible for the observation of inverse layer-dependence of friction on Re-doped MoS$_2$ samples, we performed AFM-based roughness measurements in contact mode. In particular, Fig. S1(a) shows roughness measurements on undoped MoS$_2$. While mean roughness values recorded on 1-, 2- and 3-layer regions are nearly identical, the mean roughness of the bulk region is ~11% lower, in accordance with previous studies that proposed reduced roughness at increasing number of layers as an alternative / complementary mechanism to the puckering effect [3]. On the other hand, the roughness measurements performed on Re-doped MoS$_2$ (reported in Fig. S1(b)) show no significant change between different regions, with the difference between mean roughness values on different number of layers being less than ~2.2%. These findings demonstrate that changing roughness with increasing number of layers cannot be the reason behind the observation of anomalous layer-dependence of friction on Re-doped MoS$_2$ samples.

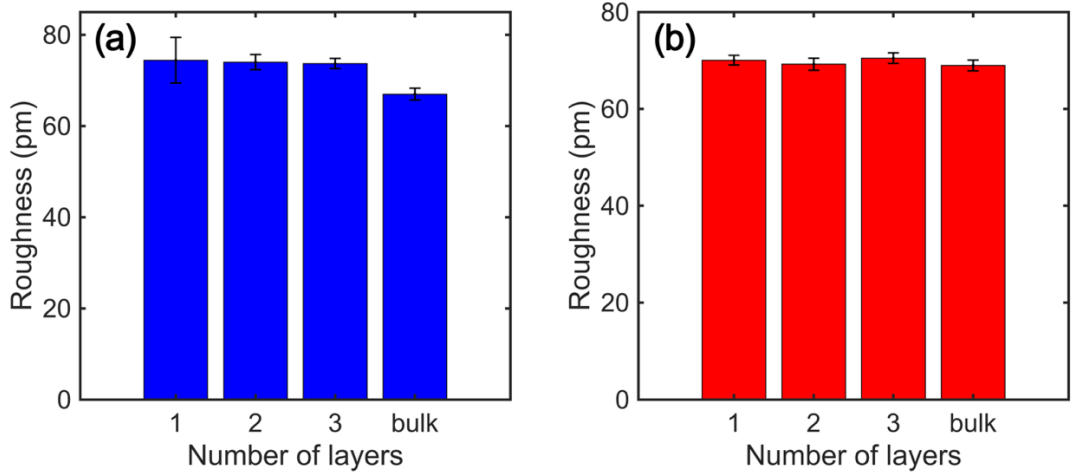

**FIG. S1.** RMS roughness values measured on 1-, 2-, 3-layer and bulk regions of (a) undoped MoS$_2$, and (b) Re-doped MoS$_2$ extracted from 10 nm × 10 nm scans.

*1.5 Adhesion Measurements*

An alternative physical mechanism potentially responsible for the observation of anomalous layer-dependence of friction on Re-doped MoS$_2$ samples could involve increasing adhesion to the AFM tip (and thus friction) at increasing number of layers [4]. In order to probe whether there is any effect of number of layers on adhesion force, we performed force spectroscopy experiments on undoped and Re-doped MoS$_2$ samples to extract adhesion values. In particular, adhesion maps consisting of 256 × 256 pixels were recorded on the samples areas presented in Fig. 1(a) and Fig. 1(c) to extract adhesion values with high statistical significance. As demonstrated in Fig. S2, no significant difference in adhesion forces exist between single-layer and few-layer regions of the flakes for both undoped and Re-doped MoS$_2$ samples, as previously reported for a number of 2D materials [5]. Consequently, adhesion trends cannot explain the anomalous friction trend observed on Re-doped MoS$_2$.

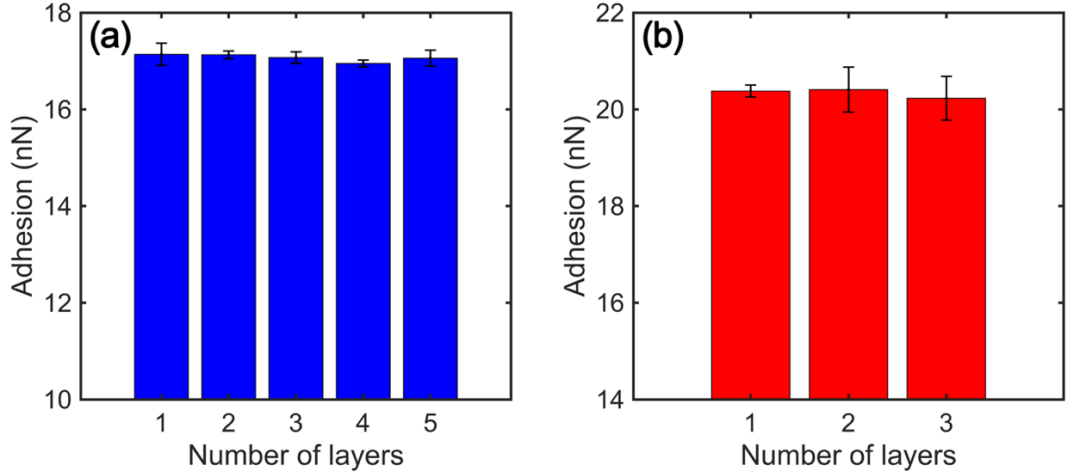

**FIG. S2.** Adhesion force values measured on (a) 1-, 2-, 3-, 4- and 5-layer regions of undoped $MoS_2$, and (b) 1-, 2-, and 3-layer regions of Re-doped $MoS_2$.

*1.6 Comparison of Friction between Doped and Undoped MoS$_2$*

**Table S1.** Mean friction values recorded on an undoped $MoS_2$ flake and two Re-doped $MoS_2$ flakes as a function of number of layers, normalized with respect to the friction recorded on the $SiO_2$ substrate around the flakes.

| **Undoped MoS$_2$** | 1-layer | 5-layer |
|---|---|---|
| | 0.65 | 0.51 |

| **Re-doped MoS$_2$ (Flake 1)** | 1-layer | 2-layer | 3-layer |
|---|---|---|---|
| | 0.37 | 0.40 | 0.50 |

| **Re-doped MoS$_2$ (Flake 2)** | 2-layer | 11-layer | 13-layer | 14-layer | 15-layer |
|---|---|---|---|---|---|
| | 0.43 | 0.49 | 0.49 | 0.55 | 0.53 |

## 2. Computational Methods and Results

*2.1 Density functional theory parameters*

We used plane-wave density functional theory (DFT) and density functional perturbation theory (DFPT) [6] implemented in Quantum ESPRESSO version 6.6 [7, 8]. Calculations were performed using the Perdew-Burke-Ernzerhof (PBE) generalized gradient approximation [9] with the Grimme-D2 (GD2) Van der Waals correction [10] and Perdew-Wang local density approximation (LDA) [11]. The pseudopotentials used were obtained from PseudoDojo [12]. LDA was used for vibrational calculations, with 80 Ry plane-wave cutoff, while PBE+GD2 was used otherwise, with 60 Ry plane-wave cutoff, as in our study of Ni-doped $MoS_2$ [13]. We worked with periodic systems of the bulk 2H structure, and considered Mo-substitution, S-substitution, tetrahedral (t-) intercalation, and octahedral (o-) intercalation, as likely stable structures by analogy to Ni-doped $MoS_2$ [13]. We found Mo-substitution and t-intercalation are the most reasonable structures to consider further. Further details on methods and analysis of our calculations are presented in a forthcoming paper [14].

*2.2 Raman spectrum calculations*

Since the Re-doped structures were *n*-type (as in experiment [15]) and therefore treated as metallic within the code, their Raman tensors are not calculable within the standard zero-frequency DFPT approach [16] because there is no gap. However, physically, the Raman tensors at optical frequencies should be only weakly affected by the small change in electron density from the dopants. As detailed in our forthcoming work [14], we developed an approximate calculation approach based on this physical picture to circumvent the limitations of the DFPT implementation. We approximated the atomic Raman tensors for Mo and S to be the same as the pristine case, and combined these with the phonon frequencies and displacement patterns calculated by DFPT for the doped system. We benchmarked this approach for (non-metallic) Ni-doped MoS$_2$ structures we have previously calculated [13] and found good results. The Re atomic Raman tensor is set to be the same as Mo in the Mo-substituted case, and is set to our intercalated Ni result [13] in the intercalated case (assuming similar behavior for different intercalated transitional metals in MoS$_2$). The results are plotted in Fig. 4.

*2.3 Elasticity calculations*

$C_{33}$ elastic parameters were computed using the stress-strain relationship (using Voigt notation) $\sigma_i = C_{ij}\epsilon_j$. Bulk pristine, Mo-substituted, and tetrahedral intercalated MoS$_2$ (of $2 \times 2 \times 1$, $2 \times 2 \times 2$, $3 \times 3 \times 1$, and $4 \times 4 \times 1$ supercells, in order of decreasing Re concentration) were subjected to uniaxial strains ranging from -0.002 to 0.002 in intervals of 0.001 in the out-of-plane direction. Monkhorst-Pack k-grids for each supercell were respectively $3 \times 3 \times 2$, $3 \times 3 \times 1$, $3 \times 3 \times 2$, and $2 \times 2 \times 2$. Each structure's atoms were relaxed for each strain while holding the lattice vectors constant. $C_{33}$ was determined by linear regression on the calculated stress vs. strain. Results are plotted in Fig. 5.

*2.4 AFM friction force calculations*

The AFM friction force calculations were carried out using a model Si cluster which has one Si at the tip and is passivated with H atoms on the top, to simulate the AFM tip apex, as used previously in the literature [17]. A *k*-grid of 3×3×1 was used. The tip was fixed 3.5 Å above the S plane. We rigidly translated the tip across a 4×4 slab of MoS$_2$ in a direction passing over successive Mo atoms, two rows away from the Re dopant. This choice, for computational convenience, is expected to over-emphasize the dopant effect since the AFM tip would typically be farther away from the dopant laterally. We determine energy at each step of $\Delta x$ = 0.32 Å, and compute forces by finite differences along the sliding direction, which we plot as a histogram. Layers were added to the MoS$_2$ slab while keeping the bottom of the MoS$_2$ and the AFM tip separated by an adequate constant vacuum of 12 Å. We found that which layer the Re atom was in had negligible effect on the force distribution when placing the Re atom more than 2 layers below the tip. We also found that adding layers beyond the Re dopant did not affect the force distribution (i.e. the 3-layer and 4-layer structure with Re in between layers 2 and 3 in both cases result in the same force distribution). We use this observation to plot the force distributions (Fig. S3) as a function of the number of layers, averaging over the possible Re locations in each layer. We see that forces are generally reduced as the number of layers increases, which is opposite to the measured trend in AFM friction. This fact indicates that the mechanism of the inverse layer dependence is not the direct interaction of the AFM tip with a rigid flat MoS$_2$ surface, but instead has to do with elastic effects on puckering. Our forces for many-layer Re-doped structures saturate at values significantly larger than those for pristine, which seems to be due to an electronic polarization effect which may be an artefact of the DFT functional approximations. We note that the relation between these calculated force distributions and measured friction forces is not a simple one, but based on the concept of stick-slip motion, we may expect that measured friction is related primarily to the larger-force end of the distribution [18].

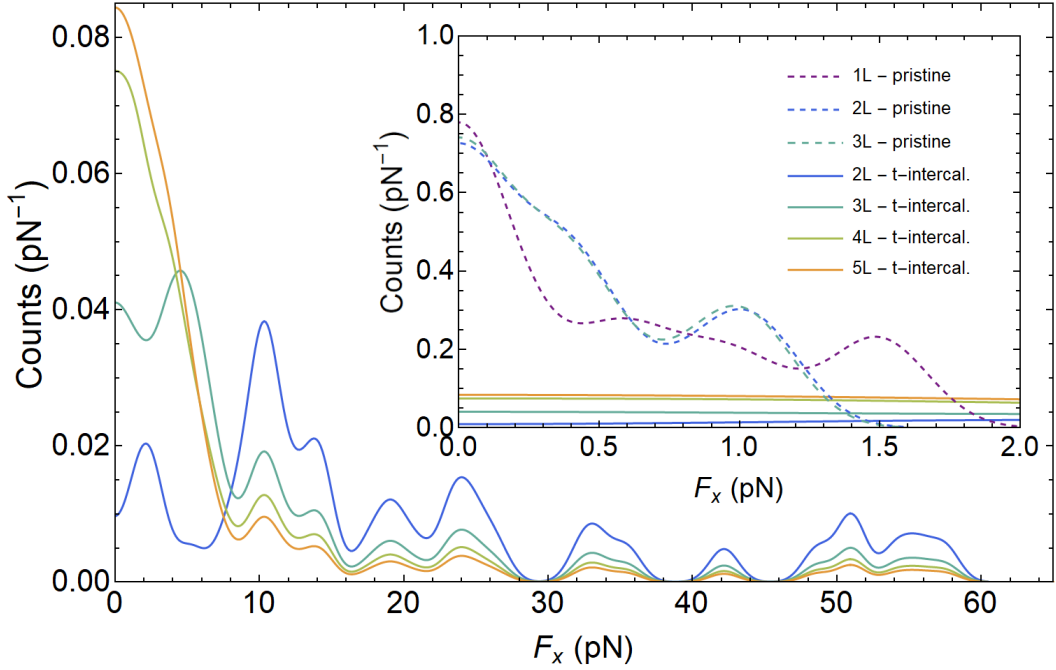

**FIG. S3.** Friction force distributions for t-intercalated Re-doped $MoS_2$ with different numbers of layers, compared with pristine $MoS_2$, calculated via DFT. Each distribution describes the friction force experienced during sliding of the model AFM tip apex over an *N*-layer system, averaging over possible locations of the Re dopant among the layers. For the doped systems, the average magnitude of the forces decreases with number of layers, and then saturates around 4 layers. A smoothing parameter of 1 pN is used for plotting with Mathematica's *SmoothHistogram* function [19]. Inset: Expanded view to compare pristine $MoS_2$, having much lower forces, with doped $MoS_2$; two and three layers are almost identical. Smoothing parameter is 0.175 pN.